\documentclass[sigconf]{acmart}
\usepackage{graphicx}
\usepackage{array}
\usepackage{float}
\usepackage{booktabs}
\usepackage{makecell}
\usepackage{multirow}
\usepackage{geometry}
\usepackage[htt]{hyphenat}
\usepackage{subcaption}
\usepackage{enumitem}
\usepackage{float}
\usepackage[most]{tcolorbox}
\usepackage{threeparttable}
\usepackage{soul}

\newcolumntype{L}[1]{>{\raggedright\arraybackslash}p{#1}}
\newcolumntype{C}[1]{>{\centering\arraybackslash}p{#1}}
\newcommand{\etal}{\textit{et al.}\ }

\begin{document}

\author{Yunsen Lei}
\email{yunsen.lei@gwu.edu}
\affiliation{%
  \institution{The George Washington University}
  \city{Washington}
  \state{DC}
  \country{USA}}

\author{Kexin Bai}
\email{baikexin@tamu.edu}
\affiliation{%
  \institution{Texas A\&M University}
  \city{College Station}
  \state{TX}
  \country{USA}}

\author{Quan Li}
\email{quanli@tamu.edu}
\email{quan.li@princeton.edu}
\affiliation{%
  \institution{Texas A\&M University}
  \city{College Station}
  \state{TX}
  \country{USA}}
\affiliation{\institution{Princeton University}
\city{Princeton}
  \state{NJ}
  \country{USA}}

\author{H. Howie Huang}
\email{howie@gwu.edu}
\affiliation{%
  \institution{The George Washington University}
  \city{Washington}
  \state{DC}
  \country{USA}}

\begin{abstract}
Export controls have become one of America's most prominent tools of economic statecraft. They aim to block rival countries' access to sensitive technologies, safeguard U.S. supply chains, protect national security, and shape geopolitical competition. Among various instruments, the U.S. Entity List has emerged as the most salient, yet its dynamics remain underexplored. This paper introduces a novel temporal graph framework that transforms the Entity List documents from a static registry of foreign entities of concern into a dynamic representation of geopolitical strategy. We construct the first event-based dataset of U.S. government foreign entity designations and model them as a temporal bipartite graph. Building on this representation, we develop a multi-level analytical approach that reveals shifting roles, enforcement strategy, and broader sanction ecosystems. Applied to 25 years of data, the framework uncovers dynamic patterns of escalation, persistence, and coordination that static views cannot capture. More broadly, our study demonstrates how temporal graph analysis offers systematic computational insights into the geopolitical dynamics of export controls.
\end{abstract}

\ccsdesc[500]{Information systems}
\ccsdesc[500]{Information systems~Information retrieval}
\ccsdesc[500]{Information systems~Document collection models}

\keywords{Graph Modeling, Temporal Bipartite Graph, Sanctions Network Analysis}

\title[Temporal Graph Theoretic Analysis of Geopolitical Dynamics in the US Entity List]{Temporal Graph Theoretic Analysis of Geopolitical Dynamics in the U.S. Entity List}
\maketitle

\section{Introduction}
\label{sec:intro}
The past few decades have seen a sharp rise in global strategic rivalry. Countries have recognized that advanced technologies, such as semiconductors, AI, and quantum computing, are just as crucial as traditional defense in protecting national security and achieving foreign policy objectives~\cite {miller2022chip}. As a result, governments in the United States, the European Union, Japan, China, and the United Kingdom have expanded their use of export controls~\cite{drezner2024global}. These regulations restrict the transfer of goods, software, and technological know-how to foreign entities that are considered to pose risks and threats to US national security~\cite{mastanduno1988trade, baldwin2020economic}.

Among these measures, the U.S. Entity List~\cite{entity_list} has become one of the most influential tools since its creation in 1997. The U.S. government prohibits U.S. firms from doing business with those entities on the list or requires them to obtain government licenses to do so. The U.S. government's increasing use of the Entity List has thus drawn growing attention from firms, policymakers, and scholars.  For firms, these listings have significant business consequences, directly impacting market access, supply chains, and compliance costs~\cite{li2024sanction, bai2025huawei, liu2025, hu2024, shen2024}. For policymakers, the listings reflect U.S. efforts to balance national security risks against the need for American firms to profit from the sale of goods, services, and technological know-how in an interdependent global economy without a central governing authority. For scholars, the listings provide important information about how the U.S. government employs targeted sanctions to achieve its strategic objectives. It is important to note that Entity List designations are more nuanced and targeted than traditional trade embargoes; they serve the national security and foreign policy interests of the U.S. rather than working as typical trade barriers for economic reasons.

However, despite its importance and increasing use, the Entity List has just begun to receive scholarly attention ~\cite{ney2021, pagano2023, liu2025, hu2024}. These recent analyses often focus narrowly on Chinese firms. The primary reason for the lack of a systematic understanding of the Entity List, particularly regarding the strategic dynamics in entity designations, is the absence of systematic data on the Entity List, and more specifically, the strategic intent and dynamics embedded within lengthy, dense, and unstructured Federal Register documents. 

In this paper, we aim to answer the question: \textit{How have U.S. targeted sanction strategies, as reflected in the Entity List, evolved, and what patterns emerge at the country level?} To this end, we introduce the first comprehensive framework for extracting, modeling, and analyzing the strategic dynamics of Entity List designations. The contributions of this paper are summarized as follows:

\begin{figure*}[htbp]
    \centering
    \includegraphics[width=0.95\textwidth]{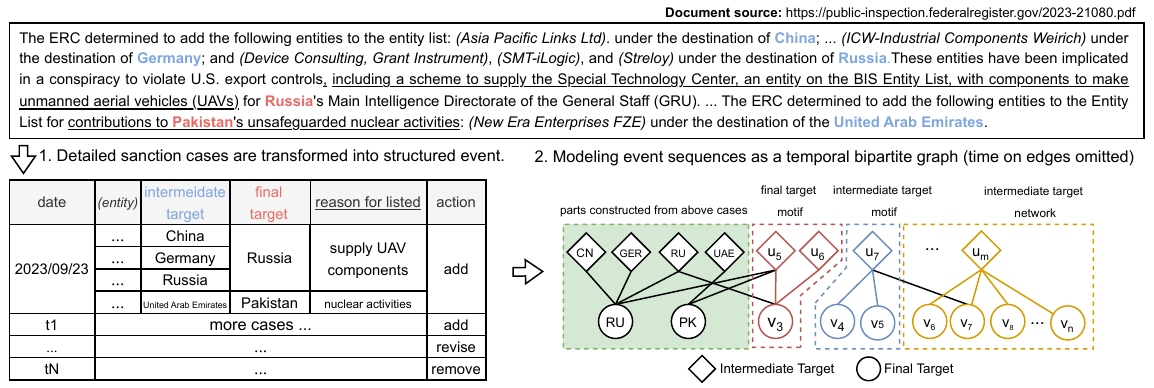}
    \caption{An overview of the data processing pipeline from (1) event extraction to (2) graph construction}
    \label{fig:process_pipeline}
\end{figure*}

\textbf{Pipeline Construction of Sanction Events}: We develop a document\hyp{}to\hyp{}event pipeline that converts over 25 years of unstructured regulatory text in Federal Register notices into a normalized event schema. The schema encodes entity identifiers, justifications, and temporal markers for event sequencing. This pipeline enables systematic reconstruction of Entity List life cycles and supports downstream graph-based analysis.

\textbf{Temporal Bipartite Graph Modeling}: We develop a temporal bipartite graph model that translates entity-level listings into state-level strategic relations, linking \textit{final target countries} (whose capabilities the Entity List ultimately seeks to restrict) to \textit{intermediate target countries }(whose entities are added solely because they do business with final targets' designated entities).

\textbf{Graph Analytical Framework}: We develop a multi-level analytical approach. At the node level, we propose the Role Skew Index (RSI) to measure whether a country is treated primarily as a transshipment hub or as a direct adversary. At the edge level, we use temporal motif analysis to uncover recurring restriction campaigns. At the subgraph level, we identify networks that reveal a broader enforcement landscape against a country. Applying this framework to 25 years of Entity List data, we uncover shifting targeted sanction strategies across presidential administrations and highlight patterns invisible in static datasets.

\section{Background on Entity List}
Sanctions are broadly defined as restrictions on the flow of goods, services, or capital to achieve foreign policy or security objectives~\cite{hufbauer1990economic}, and export controls fit this scope. U.S. export control rules regulate what items can be exported, when, where, and to whom. Like traditional sanctions, their aim is to ``reduce the power or threat posed by a target''~\cite{farmer2000costs}.  The Bureau of Industry and Security (BIS), under the U.S. Department of Commerce, describes the Entity List as identifying entities engaged in activities that pose a threat to U.S. national security or foreign policy interests. The U.S. government explicitly treats export controls as sanctions. For example, in 2014, the BIS stated that new Entity List rules were ``imposing additional sanctions'' in response to Russia’s destabilization of Ukraine~\cite{fedreg2014-18579}. Scholars also conceptualize export controls as targeted sanctions, incorporating them into datasets~\cite{morgan2014threat, portela2023evolution} and analyzing their role in modern economic statecraft~\cite{allen2021targeted}.

To control exports, re-exports, and transfers to parties of concern,  the BIS manages four key lists at the entity level: the Entity List, the Unverified List, the Denied Persons List, and the Military End-User List. This paper focuses on the Entity List because of its large size, rapid expansion, and high salience. The \textit{Entity List}~\cite{entity_list} designates foreign firms, research institutions, government bodies, and individuals as entities of national security or foreign policy concern. Consequently, exports or re-exports to these entities are either directly prohibited or require a license from the US government. 
Unlike the Unverified List~\cite{uv_list}, which only triggers heightened due diligence, the Entity List imposes binding licensing restrictions that directly block access to certain U.S. goods and technologies. The Entity List carries greater policy weight than the Denied Persons List~\cite{dp_list}, which primarily targets past violators, and the Military End-User List~\cite{meu_list}, which focuses on a narrower set of military-linked actors. The Entity List serves not only as a compliance mechanism but also as a forward-looking instrument to prevent sensitive goods and technologies from reaching foreign adversaries.

\section{Data Processing and Description}
We collect U.S. Entity List notices from the Federal Register’s public API~\cite{fr_api}, retrieving 405 documents published between 1997 and the present. Each notice follows a consistent structure: a ``Supplementary Information'' section outlining the rationale for the action, and a table listing sanctioned entities, including aliases, addresses, and licensing requirements. Our goal is to transform unstructured notices into a structured, event-based dataset for longitudinal analysis. This task requires extracting relevant information and analyzing it to reveal hidden patterns in policy restrictions.

\subsection{Intermediate Target and Final Target}
\label{sec:state_dyad}
The Entity List documents provide the sanctioned entities and the reasons for their inclusion. These justifications often reveal a dyadic relationship between two states, which is central to our analysis.  

For example, in one Federal Register notice~\cite{sanction_example} (illustrated in Figure~\ref{fig:process_pipeline}), two distinct sanction cases are recorded. The first case adds multiple entities from China, Germany, and Russia to the Entity List for supplying UAV-related components to a Russian entity already under restriction. The second case adds an entity from the United Arab Emirates for its contributions to Pakistan’s unsafeguarded nuclear activities. 

To capture relations between countries, we define two roles in each sanction case: the \textbf{final target country}, whose military or technological capabilities the U.S. government ultimately seeks to restrict, and the \textbf{intermediate target country}, whose entities are added solely because they do business with final targets' designated entities. For simplicity, we refer to them as the \textbf{final target} and the \textbf{intermediate target} throughout the paper.
 
Such a country dyad, identified at the entity level for each event, is key to understanding how the Entity List operates. While the Entity List's restrictions apply to non-state actors, the ultimate goal of these regulatory actions is to influence a foreign state's behavior, policies, or capabilities~\cite{achilleas2017introduction}. Listed entities are best understood not as the ultimate targets, but as proxies for the target countries. Separating the final target and intermediate target allows researchers to investigate the US government's detection and restrictions of the so-called ``sanction busters'', i.e., third parties that help the final target countries to evade restrictions~\cite{early2009sleeping, early2011unmasking, early2015busted, early2021making}. Demonstrating how the U.S. government blocks exports to rival end-users and those who assist them sheds light on American economic statecraft.

\subsection{Document Parsing}
\label{sec:doc_process}
Federal Register notices are documents with both structured tables and unstructured narratives. We adopt a divide-and-conquer strategy. A custom parser first extracts tabular sections, generating a list of entities with aliases and known addresses. We then build an embedding index over the notice’s chunked narrative text and, for each entity, query the index with its name plus aliases. The top-$k$ most retrieved chunks are then passed to a Large Language Model (LLM)~\cite{lewis2020retrieval} for targeted information extraction. 

Following Section~\ref{sec:state_dyad}, the LLM is prompted to extract two attributes that allow us to move from entity-level designations to state-level relations: (1) Reason for listing---the official justification, such as proliferation activity or military end-use. (2) Final target---the state whose capabilities the sanction seeks to restrict. As shown in Figure~\ref{fig:process_pipeline}, reasons for listing (e.g., “supply UAV components”) are often explicitly stated, whereas final targets require inferring implicit relationships. To align the extraction process with the correct strategic logic, our prompts provide the LLM with explicit rules and heuristics. First, if an entity is listed for supporting or enabling the activities of another state (e.g., through technology transfers or logistical support), that state is designated as the final target. Second, if an entity is listed for its own misconduct or because the U.S. considers it a direct threat to national security, then the firm’s home country is treated as the final target.

The model outputs a structured JSON record, 
merged with parsed table fields and metadata to yield a sanction event: \texttt{event = (date, \allowbreak entity, \allowbreak intermediate target, \allowbreak final target, \allowbreak reason, \allowbreak action)}. The intermediate target derives from the table parser. Parsing all notices yields a chronological sequence of sanction events that enables reconstruction of the Entity List at any historical point. The dyadic state relations embedded in these events naturally translate into a temporal bipartite graph, whose construction we describe in Section~\ref{sec:graph_build}. This structure enables us to systematically study trends in entity designations and utilize entity-level actions to identify state-level sanction networks. We include more details about the language model configuration, parsing example, and discussion about parsing errors in Appendix~\ref{sec:appendix_llm}.

\subsection{Data Description}
\begin{figure}[t]
\centering
\includegraphics[width=0.8\linewidth]{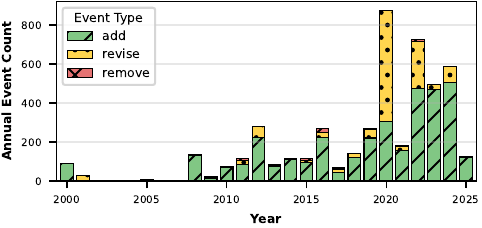}
\caption{Entity List Restriction Events and Distribution}
\label{fig:event_trend}
\end{figure}

From all documents collected, we parsed 4,808 sanction-related events. The distribution of event types is highly imbalanced, with the majority being additions (3,547), followed by revisions (1,192) and removals (69). Figure~\ref{fig:event_trend} plots the distribution of these events from 2000 to March 2025, illustrating how enforcement activity has intensified over time. In particular, the frequency of new designations rose sharply after the passage of the Export Control Reform Act (ECRA) of 2018~\cite{HR5040_ECRA2018}.

Entity List restrictions are not evenly distributed across countries. Table~\ref{tab:top_countries} lists the top ten countries involved, categorized as either intermediate or final target. In the intermediate target category, China (33.2\%) and Russia (28.3\%) together account for over 61\% of all cases. Among the final target countries, Russia (36.9\%), China (24.8\%), and Iran (15.5\%) account for more than 77\% of the total.

\begin{table}[htbp]
\centering
\begin{threeparttable}
\caption{Top ten intermediate targets and final targets}
\label{tab:top_countries}
\begin{tabular}{|r l r|r l r|}
\hline
\multicolumn{3}{|c|}{\textbf{Intermediate Target}} & 
\multicolumn{3}{c|}{\textbf{Final Target}} \\
\hline
1. & China (CN) & 33.2\% & 1. & Russia (RU) & 36.9\%  \\
2. & Russia (RU) & 28.3\%  & 2. & China (CN) & 24.8\%  \\
3. & UAE & 6.0\%  & 3. & Iran (IR) & 15.5 \\
4. & Pakistan (PK) & 5.5\%  & 4. & Pakistan (PK) & 6.1\%  \\
5. & Iran (IR) & 3.7\%  & 5. & Afghanistan (AF) & 2.8\%  \\
6. & Turkey (TR) & 2.2\%  & 6. & Ukraine (UA) & 2.4\%  \\
7. & Malaysia (MY) & 1.5\%  & 7. & Syria (SY) & 1.7\%  \\
8. & UK & 1.5\%  & 8. & Iraq (IQ) & 1.6\% \\
9. & India (IN) & 1.5\%  & 9. & India (IN) & 1.2\%  \\
10. & Singapore (SG) & 1.3\%  & 10. & UAE & 0.8\%  \\
\hline
\multicolumn{2}{|r}{\textbf{Total}} & \textbf{84.7} &
\multicolumn{2}{r}{\textbf{Total}} & \textbf{93.9} \\
\hline
\end{tabular}
\begin{tablenotes}
\small
\item[a] UK = United Kingdom; UAE = United Arab Emirates
\end{tablenotes}
\end{threeparttable}
\end{table}

We also code the reasons for entity designations based on the reasons provided. 
As shown in Table~\ref{tab:reason_area}, the top four reasons are evasion \& supply-chain facilitation (34.62\%), conventional military \& defense systems (32.62\%), WMD programs \& delivery systems (13.05\%), and semiconductors \& advanced computing (8.46\%), amounting to 88.75\% of all reasons. The ``Other/Unknown'' reason refers to justifications that contain only generic policy language or mention activities involving ``unspecified'' items or technologies. More details about the coding process is discussed in Appendix~\ref{sec:appendix_cat}.

\begin{table}[htbp]
\centering
\caption{Reasons for Entity Designations}
\label{tab:reason_area}
\begin{tabular}{lr}
\toprule
\textbf{Reason} & \textbf{Percentage} \\
\midrule
Evasion \& Supply-Chain Facilitation     & 34.62\% \\
Conventional Military \& Defense Systems & 32.62\% \\
WMD Programs \& Delivery Systems         & 13.05\% \\
Semiconductors \& Advanced Computing     &  8.46\% \\
Other/Unknown                            &  4.06\% \\
Human Rights \& Surveillance             &  3.61\% \\
Energy \& Critical Infrastructure        &  2.48\% \\
Cyber \& Secure Communications           &  1.10\% \\
\midrule
\textbf{Total}                           & 100.00\% \\
\bottomrule
\end{tabular}
\end{table}

\section{Methodologies for Analyzing Entity List Restrictions}
\label{sec:framework}
The event-based dataset described above provides a rich, temporal record of U.S. export control actions. Using the new dataset, we examine three research questions derived from our overarching question.

\textbf{RQ1}: \textit{What roles do countries play as intermediate or final targets in U.S. sanctions, and how do these roles reflect shifts in U.S. foreign policy?}

\textbf{RQ2}: \textit{How do recurring temporal motifs in the Entity List reveal the sequencing and intensity of U.S. sanction campaigns?}

\textbf{RQ3}: \textit{How do the structures of intermediate and final target networks differ across major U.S. rivals, and what do these patterns reveal about U.S. enforcement challenges?}

To this end, we introduce a multi-level analytical framework based on a temporal bipartite graph representation. This formulation allows us to analyze U.S. sanction strategies at three levels: node (how countries are perceived), edge (how actions are sequenced), and subgraph (how broader structures emerge).

\subsection{Graph Formulation}
\label{sec:graph_build}

Our processing pipeline (Section~\ref{sec:doc_process}) yields a chronological sequence of sanction events. Each event is represented as: $event = (date,\allowbreak entity,\allowbreak c^{int},\allowbreak c^{fin},\allowbreak reason,\allowbreak action)$, where $c^{int}$ is the entity’s intermediate target, $c^{fin}$ is the final target, and $action\in \{add, revise,  \allowbreak remove\}$. We model these event-level data into a temporal bipartite graph $G = (V, E)$:

\textbf{Nodes ($V$)}: The node set is partitioned into intermediate target $C_{int}$ and final target $C_{fin}$. These sets are disjoint in the graph representation, i.e., $C_{int} \cap C_{fin} = \emptyset$. A country $c$ that plays both roles is represented by two distinct nodes, $c^{int} \in C_{int}$ and $c^{fin} \in C_{fin}$, which correspond to the same underlying country but are distinguished by role.

\textbf{Temporal Edges ($E$)}: Each sanction case defines a directed temporal edge $e_k = (u_k, v_k, t^{add}_k, t^{remove}_k)$, where $u_k \in C_{int}$ and $v_k \in C_{fin}$. $t^{add}_k$ is the date of addition, and $t^{remove}_k$ is the removal date (or $+\infty$ if still active). This edge represents the full life-cycle of the sanction against entity $k$.

For clarity, we use projection functions $u(\cdot)$, $v(\cdot)$, $t^{add}(\cdot)$ and $t^{remove}(\cdot)$ to access an edge's corresponding attributes. This representation elevates entity-level actions into a country-level temporal network, enabling analysis of sanction strategies over time.

\subsection{Node-Level: The Role Skew Index}
\label{sec:rsi_def}
In the Entity List, as shown by the sanction case in Figure~\ref{fig:process_pipeline}, a country can assume two roles. As an intermediate target, it functions more as a transshipment hub or re-exporter. The country is an indirect target leveraged through enforcement. As a final target, it is considered an adversary that poses threats to the national security or foreign policy interests of the United States.

To quantify this duality, we define the \textit{Role Skew Index} (RSI) for country $c$ at time $t$:

\begin{equation}
    RSI(c,t) = \frac{deg(c^{int}, t) - deg(c^{fin}, t)}{deg(c^{int}, t) + deg(c^{fin},t)}
\end{equation}
where $\deg(c^{int}, t)$ is the degree of $c$ as an intermediate target, and $\deg(c^{fin}, t)$ is its degree as a final target, both measured over the set of active edges at $t$.
When $RSI(c,t) = 1$, country $c$ is detected as an intermediate target in all cases; when $RSI(c,t) = -1$, it is a final target in all cases; when $-1 < RSI(c, t) < 1$, country $c$ is considered to have different roles in different cases.

Tracking RSI over time can reveal substantive role shifts. A shift from an intermediate target to a final target indicates a geopolitical escalation. Conversely, a shift from a final target to an intermediate target suggests that the country is increasingly acting as a re-export hub rather than being the primary target.

\subsection{Edge-Level: Motif as Restriction Intensity}
\label{sec:motif_def}
While RSI provides a valuable, high-level view of a country's role, it aggregates discrete events in a period $t$ into a single score. As a result, RSI can signal that a country's role is shifting but cannot illuminate the strategic context behind that shift. To understand sanctions as instruments of statecraft, it is crucial to recognize that they rarely occur as isolated actions. Instead, restriction events are often interdependent steps in broader campaigns aimed at disrupting evasion networks or constraining entire technology ecosystems.

\begin{figure}[htbp]
\centering
\includegraphics[width=0.9\linewidth]{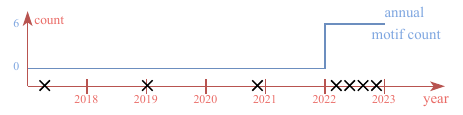}
\caption{Motif's Design Rationale}
\label{fig:motif_illustrate}
\end{figure}

As illustrated in Figure~\ref{fig:motif_illustrate}, consider a country that experienced three scattered sanction events (denoted by the asterisk) between 2017–2021, followed by four additional sanctions arriving in rapid succession in 2022. Although the total number of sanction events increased only moderately, their temporal distribution shows distinct patterns. A clearly visible intensive sanction campaign, as indicated by temporal clustering, occurred in 2022. To detect such a sanction campaign and quantitatively characterize the campaign-level intensity, we need a measurement that filters out isolated actions while amplifying periods of concentrated sanction activity, making episodes of strategic focus stand out clearly. 

\textbf{Definition}: We turn to an edge-level analysis using temporal motifs. Following Paranjape~\etal~\cite{paranjape2017motifs}, a $l$-edge, $\delta$-temporal motif is an ordered set of temporal edges ${e_1, e_2, \dots, e_l}$ such that the edge start times are non-decreasing, $t^{add}_1 \leq t^{add}_2 \leq t^{add}_3 \leq ... \leq t^{add}_l$, and all edges occur within a bounded time window. Specifically, the motif satisfies $t^{add}_l - t^{add}_1 \leq \delta$, meaning the time between the first and last edge is at most $\delta$. Our bipartite graph consists of two types of motifs as illustrated by visual examples in Figure~\ref{fig:process_pipeline}: the \textit{intermediate target motif}, where two events originate from the same intermediate target, and the \textit{final target motif}, where two events converge on the same final target.

\begin{align}
\mathcal{ITM}(c^{int}; \delta) &= 
\{ e_1, e_2 \in E \ \mid \ u(e_1) = u(e_2) = c^{int} \} 
\label{eq:M_int} \\ 
\mathcal{FTM}(c^{fin}; \delta) &= 
\{ e_1, e_2 \in E \ \mid \ v(e_1) = v(e_2) = c^{fin} \} \label{eq:M_fin} 
\end{align}

A motif is valid at observation time $t$ only if (1) two sanction events, $e_1$ and $e_2$,  are in force at $t$, i.e., $t^{add}(e_1) \leq t < t^{remove}(e_1)$ and $t^{add}(e_2) \leq t < t^{remove}(e_2)$, and
(2) the two sanction events are initiated close in time, $|t^{add}(e_1) - t^{add}(e_2)| \leq \delta$.

\textbf{Interpretation}:
A motif count is the sum of weighted events. For each sanction event represented as an edge $e$, its weight equals the number of prior events targeting or originating from the same country $c$ within the window $\delta$. 

Using \textit{final target motif} as an example, $w(e) = \bigl|\{ e' \in E \ \mid\ v(e') = v(e) = c^{fin},\ 0 < t^{add}(e) - t^{add}(e') \leq \delta \}\bigr|$. The corresponding motif count in a given time bin $T$ is then the sum of these weights: $\sum_{e \in E(T)} w(e)$. This formulation means that a sanction event carries greater significance when it forms part of a concentrated sequence rather than when it stands alone. In other words, it captures the logic of escalation. Each additional event within the cluster amplifies the importance of those that follow. When applied to the example in Figure~\ref{fig:motif_illustrate} with $\delta \leq 1$ year and $T$ set to one year, the motif count remains zero before 2022 despite multiple temporally scattered events. In 2022, their weights accumulate as $0+1+2+3$, yielding a motif count of $6$.

\subsection{Subgraph-Level: Detection Networks}
\label{sec:net_def}

The node- and edge-level analyses build on roles (RSI) and recurring patterns (motifs) that focus on individual countries. To capture the broader enforcement landscape, we extend our analysis to the subgraph level. A subgraph, in our context, represents all sanction events that connect a group of countries through shared intermediate or final targets within a given time window. Figure~\ref{fig:process_pipeline} shows an example of such a subgraph when the shared node is an intermediate target.

Formally, let $t$ be the observation time. The active edge set is $E(t) = \{\, e \in E \mid t^{add}(e) \leq t < t^{remove}(e) \,\}$. For a country $c$, we define two types of detection networks, depending on whether $c$ appears as an intermediate or final target:

\begin{align}
\mathcal{ITN}(c^{int}, t) &= \{\, v(e) \mid e \in E(t), \ u(e) = c^{int} \,\} \\
\mathcal{FTN}(c^{fin}, t) &= \{\, u(e) \mid e \in E(t), \ v(e) = c^{fin} \,\} 
\end{align}

We call $\mathcal{ITN}(c^{int}, t)$ an \textit{Intermediate Target Network} because it's a set of countries linked to the same intermediate target $c^{int}$ by active sanctions at time $t$. Similarly, we call $\mathcal{FTN}(c^{fin}, t)$ a \textit{Final Target Network}.

These definitions extend the motifs to the subgraph scale. Whereas motif counts provide a quantitative measure for the intensity of prominent sanction campaigns, network structures reveal the broader constellation of countries involved and the relationships to the targeted countries. The subgraph-level analysis maps how multiple sanction events that are simultaneously in force during a given period interconnect across countries, offering a comprehensive view of which states are implicated in ongoing enforcement campaigns.

\begin{figure}[t]
  \centering
  \begin{subfigure}[t]{0.95\linewidth}
    \centering
    \includegraphics[width=\linewidth]{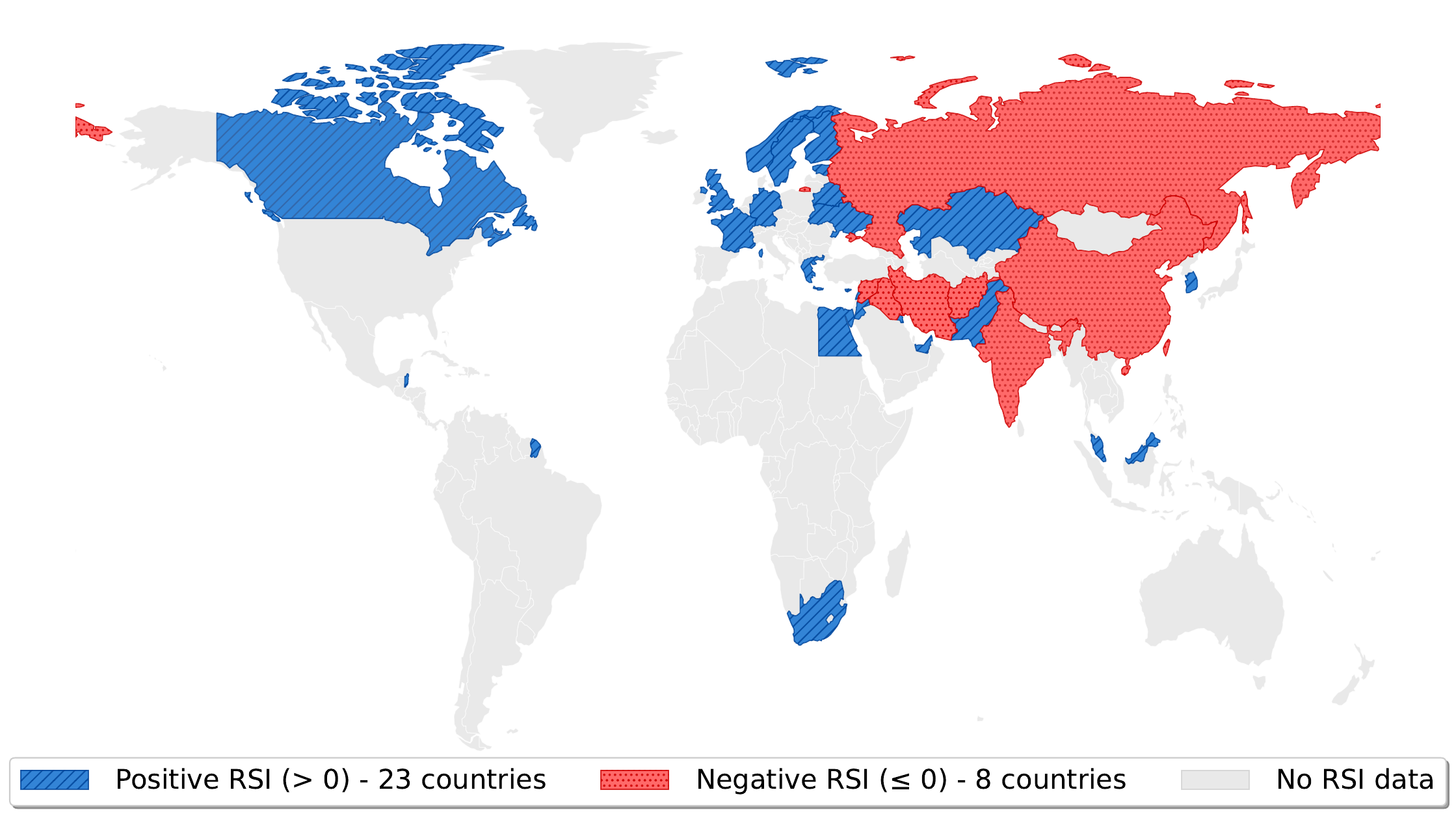}
    \caption{Cases up to 2012}
    \label{fig:rsi_world_2012}
  \end{subfigure}
  \begin{subfigure}[t]{0.95\linewidth}
    \centering
    \includegraphics[width=\linewidth]{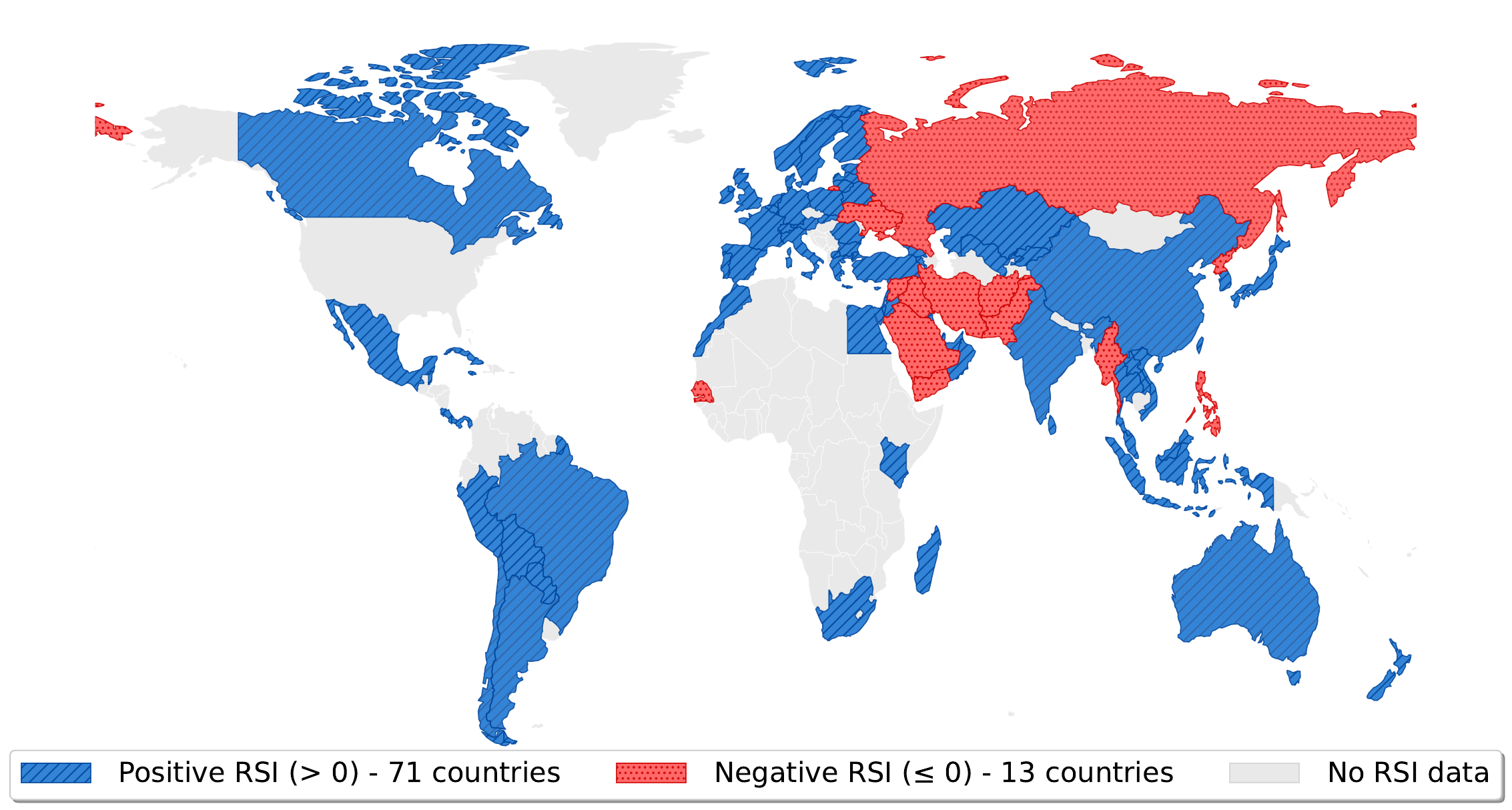}
    \caption{As of March 2025}
    \label{fig:rsi_world_2025}
  \end{subfigure}
  \caption{RSI by Countries in (a) 2012 vs (b) 2025. A country with more intermediate (or final) target events is represented by blue (positive) or red (negative) RSI values.}
  \label{fig:rsi_world}
\end{figure}

\begin{figure*}[htbp]
\centering
\includegraphics[width=\linewidth]{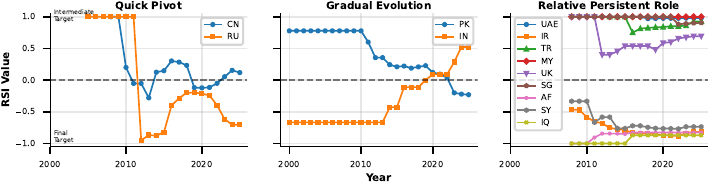}
\caption{Three Patterns of Evolution in the RSI from 2000 to March 2025 }
\label{fig:rsi_series}
\end{figure*}

\section{Findings}
\label{sec:analysis}
Applying our methodologies in Section~\ref{sec:framework} to 25 years of Entity List data, we present findings that address our research questions at three different levels of analysis below.

\subsection{A Global View of Strategic Roles}
\label{sec:rsi_analysis}
Applying our node-level metric, the Role Skew Index (RSI), to the full dataset reveals patterns in the strategic roles of different countries within the U.S. export control framework.

As shown in Figure~\ref{fig:event_trend}, Entity List designations became relatively dense beginning around 2008, reaching the first peak in 2012. To capture the distribution of country roles over time, we present two snapshots: the early stage in 2012 (Figure~\ref{fig:rsi_world_2012}) and the current landscape in 2025 (Figure~\ref{fig:rsi_world_2025}).  In 2012, only 37 countries were involved. By 2025, the number had risen to 84, showing a clear geopolitical shift. In 2025, nations typically considered U.S. adversaries, such as Iran, exhibit negative RSI values, confirming their primary status as the final targets of sanctions. Conversely, U.S. allies, including NATO and Five Eyes members, exhibit positive RSI values, indicating they are viewed as intermediate rather than ultimate targets. Notably, the RSI values of some countries shift in sign during this period, most prominently those of India and China.

China's case, however, presents a more complex picture. Despite being a primary strategic competitor, its RSI in 2025 is 0.11, indicating that more cases treat China as an intermediate target. This can be attributed to China's immense industrial capacity and deep integration into global supply chains. It has a proportionally larger number of entities that can serve as intermediate conduits for technology, both domestically and internationally.

While static analysis is insightful, tracking the RSI over time shows the dynamic evolution of U.S. sanction strategies. As plotted in Figure~\ref{fig:rsi_series}, we identify three distinct temporal patterns from the RSI time series for countries in Table~\ref{tab:top_countries}:

\textbf{Quick Pivot}: This pattern reflects an abrupt strategic shift, characterized by a sharp change in a country's RSI. For example, both Russia and China maintained an RSI of 1.0 (as an intermediate target only) for years before their values declined sharply, signaling a fundamental change in U.S. policy toward them.

\textbf{Gradual Evolution}: This pattern involves a slow, steady change over several years, reflecting an incremental evolution in U.S. strategic perceptions. Pakistan gradually shifted from an intermediate target only to a relatively balanced role, whereas India shifted from a final target to an intermediate one.

\textbf{Relative Persistent Role}: For most other top Entity List countries, their roles have remained relatively stable over time. Singapore and the UAE, for example, have been consistently intermediate targets, while Iran and Syria have been consistently final targets.

\textbf{Conclusion 1}: The RSI is an effective indicator of geopolitical alignment, particularly when a country's designation pattern is skewed toward a single role. For major economies like China, with high volumes of both intermediate and final target designations, a near-zero RSI indicates a complex dual-role status. When tracked over time, the RSI reveals underlying policy dynamics—such as gradual escalations, abrupt shifts, or persistent roles—that are entirely lost in static views.

\begin{figure*}[htbp]
\centering
\includegraphics[width=0.95\linewidth]{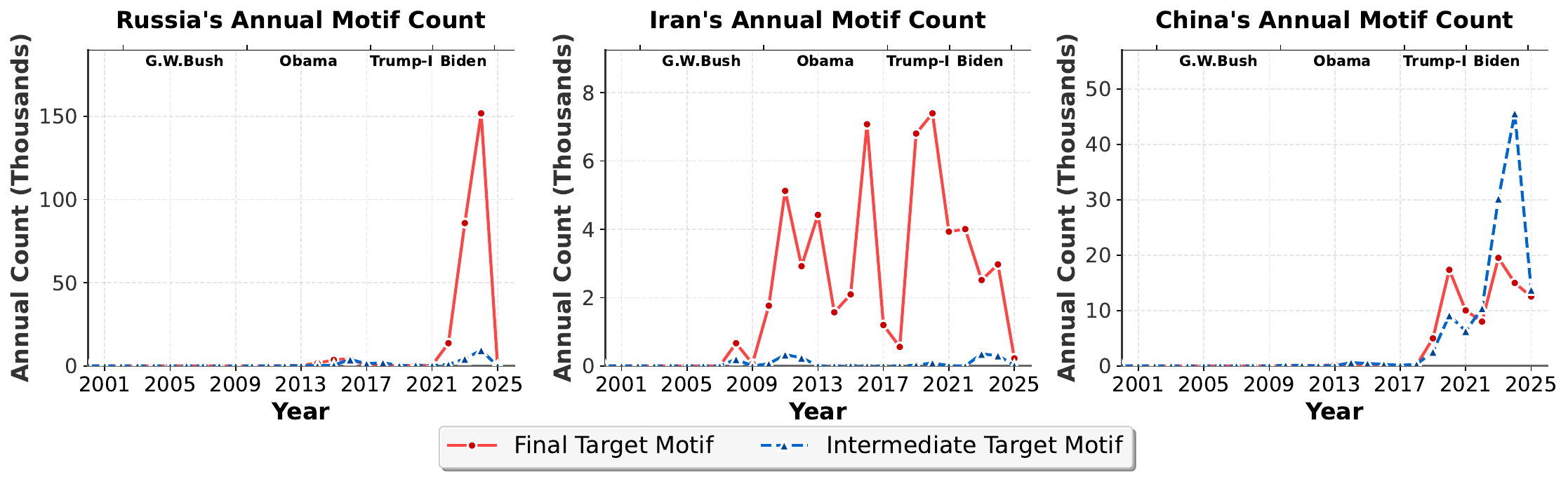}
\caption{A Comparison of Motif Counts for Russia, Iran, and China Showing Diverse Sanction Strategies}
\label{fig:three_county_motif_groups}  
\end{figure*}

\subsection{Detecting Sanction Strategies with Motifs}
\label{sec:motif_analysis}
To analyze sanction campaigns, we present in Figure~\ref{fig:three_county_motif_groups} temporal motifs within a four-year window ($\delta \leq 4$ years), corresponding to the length of a U.S. presidential term. We focus on US sanctions strategies toward the three most salient final target countries: Russia, China, and Iran. Since U.S. foreign policy is strongly shaped by presidential administrations, each with its own goals, we compute the motif counts using a one-year time bin. We observe a pattern of presidential cycles across all three final targets, with highly distinct, tailored strategies for each.

\textbf{Presidential Cycle}: 
As shown in Figure~\ref{fig:three_county_motif_groups}, a consistent pattern across all three top final targets is that sanction intensity follows a presidential cycle. Motif counts are typically low in the first year of an administration, increase progressively through the second and third years, and peak toward the end of the term. This recurring pattern reflects the operational realities of implementing foreign policy. Every presidential transition brings a new cabinet, a new national security team, and shifts in enforcement priorities. It also takes time for the new administration to determine how it will utilize the Entity List as a strategic tool.

\textbf{Russia--Reactive Escalation}: Entity List restriction campaigns toward Russia are defined by a singular, overwhelming surge in sanction intensity against Russia as a final target immediately following its full-scale invasion of Ukraine in 2022. This motif count spike stands out far more distinctly than a value drop in the RSI curve. Before 2022, Russia's motif count was minimal, indicating that Russia was not the subject of a sustained campaign. The massive spike, almost exclusively in the final target motif during Biden's administration (2021-2025), demonstrates the concentration of policy events designed to directly curtail Russian access to U.S. goods and technologies.

\textbf{Iran--Persistent Pressures}: Entity List restriction campaigns toward Iran reveal a pattern of sustained pressure against a final target over time. Significant activities began as early as 2010 and continued in cyclical waves across multiple administrations. Restriction campaigns against Iran as a final target dominated those against Iran as an intermediate target. This pattern is consistent with the long-standing U.S. policy to restrict Iran's procurement of sensitive technologies due to nuclear proliferation concerns.

\textbf{China--Strategic Pivot}: Entity List restriction campaigns toward China show a sharp strategic inflection point around 2018, coinciding with the U.S.-China trade war. Unlike Russia and Iran, China features a high volume of restriction campaigns as both intermediate and final targets. More uniquely, China's dominant motif type switched between administrations. Under the Trump administration (2017-2021), the final target motif reached its peak. In contrast, the Biden administration maintained the same level of final target restrictions but dramatically strengthened its intermediate target restrictions. The surge in the dominance of the intermediate target motif signals a strategic pivot driven by China's role as a hub in global supply chains.

The significance of a motif can be quantified through its $z$-score relative to an ensemble of random graphs, establishing whether the pattern represents a genuine organizational principle rather than random connectivity~\cite{doi:10.1126/science.298.5594.824}. We conduct further studies to demonstrate such statistical significance in Appendix~\ref{sec:appendix_null}.

\textbf{Conclusion 2}: Temporal motif analysis, which detects recurring temporal sequences of edges, provides a sharper lens on sanction dynamics than aggregate indices like the RSI. By filtering out sparse activities and amplifying bursts of targeted designations, the motif count reveals both the intensity and the sequencing of enforcement campaigns. Applied to Russia, Iran, and China, the analysis reveals three distinct patterns: reactive escalation, persistent pressures, and strategic pivot.

\subsection{Mapping Sanction Networks}
\label{sec:net_analysis}
\begin{figure*}[t]
  \centering

  \begin{subfigure}[t]{0.32\textwidth}
    \centering
    \includegraphics[width=\linewidth]{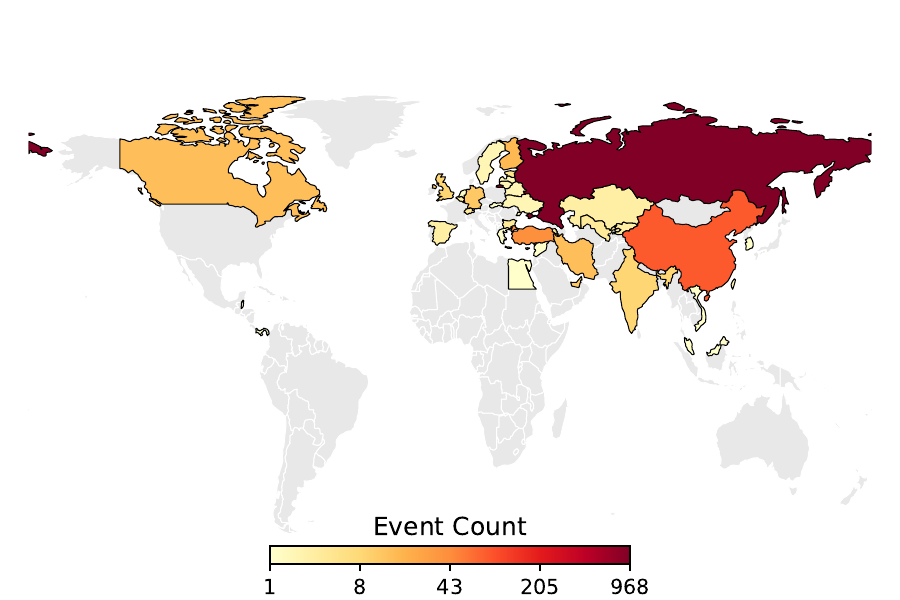}
    \caption{Russia as Final Target}
    \label{fig:russia-proc}
  \end{subfigure}\hfill
  \begin{subfigure}[t]{0.32\textwidth}
    \centering
    \includegraphics[width=\linewidth]{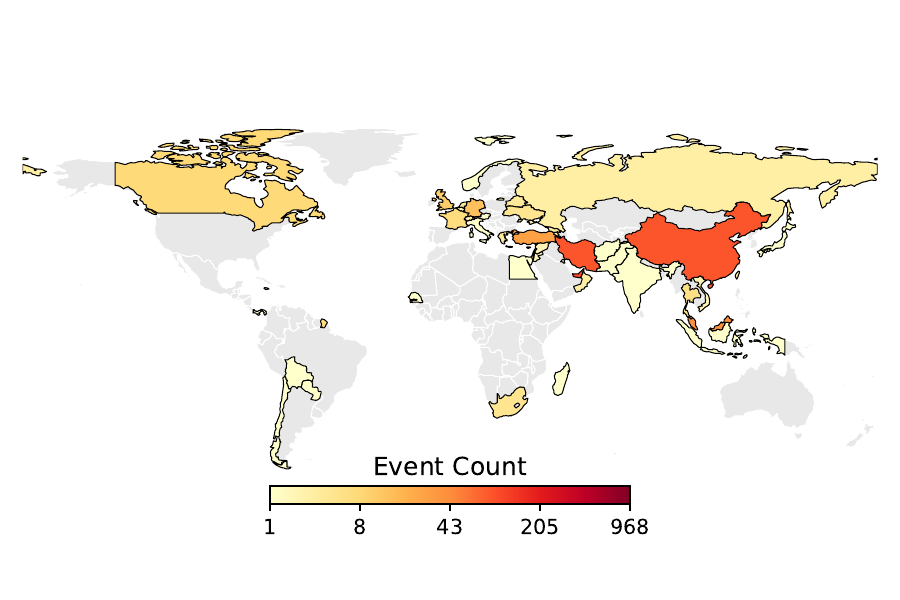}
    \caption{Iran as Final Target}
    \label{fig:iran-proc}
  \end{subfigure}\hfill
  \begin{subfigure}[t]{0.32\textwidth}
    \centering
    \includegraphics[width=\linewidth]{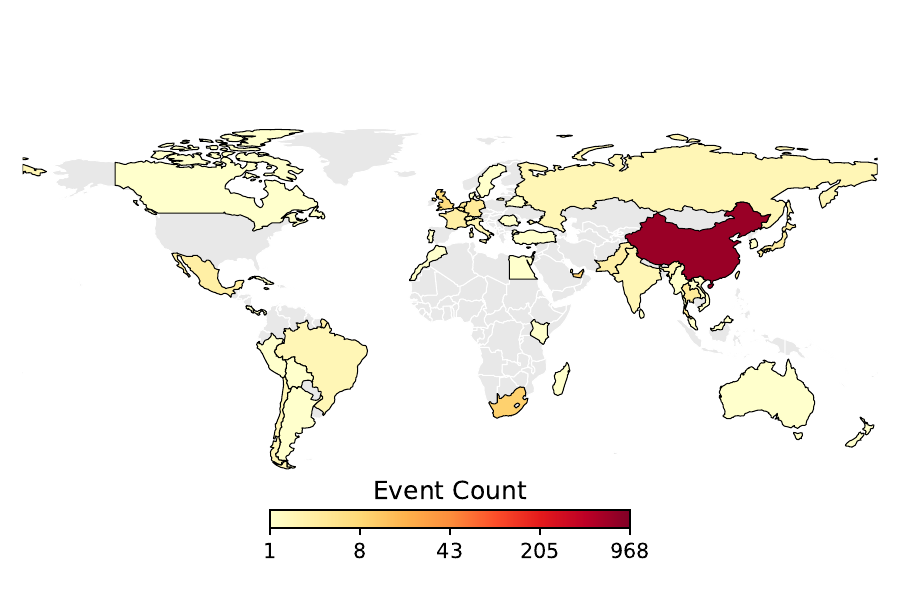}
    \caption{China as Final Target}
    \label{fig:china-proc}
  \end{subfigure}

  \caption{Final Target Networks: Russia, Iran, and China as Final Targets, cases up to March 2025.}
  \label{fig:final-networks}
\end{figure*}

We now move from individual actors and event sequences to broader network structures. Using the definitions in Section~\ref{sec:net_def}, we construct \textit{Intermediate Target Networks} (INs) and \textit{Final Target Networks} (FNs) for Russia, Iran, and China. We include an example that list each member country in a network in Appendix~\ref{sec:appendix_net}. In Figures~\ref{fig:intermediate-networks} and~\ref{fig:final-networks}, each subfigure shows the countries linked to a single target country. Countries are shaded by frequency of appearance, with darker colors indicating a higher number of sanction events within the corresponding network. Three key insights emerge.

\begin{table}[htbp]
\centering
\caption{Top 3 Members in the Networks of Top 3 Final Targets}
\label{tab:top3_networks}
\begin{tabular}{@{}p{0.18\linewidth}p{0.36\linewidth}p{0.36\linewidth}@{}}
\toprule
\textbf{\makecell{Target\\ Country}} & \textbf{\makecell{Intermediate Target \\ Network}} & \textbf{\makecell{Final Target \\ Network}} \\ 
\midrule
\makecell[c]{China}  & \makecell[c]{China (694)\\Iran (87)\\Russia (83) } 
       & \makecell[c]{China (694)\\Hong Kong (75)\\Singapore (12)} \\\hline
       
\makecell[c]{Iran}   & \makecell[c]{Iran (87)\\Russia (15)\\Iraq (11)} 
       & \makecell[c]{UAE (100) \\Iran (87)\\China (87)} \\\hline
\makecell[c]{Russia} & \makecell[c]{Russia (969)\\Myanmar (4)\\China (3) Iran (3)} 
       & \makecell[c]{Russia (969)\\Hong Kong (86) \\China (83)} \\
\bottomrule
\end{tabular}
\begin{tablenotes}
\small
\item[b] Parentheses include the number of sanction cases involving the country
\end{tablenotes}
\end{table}

\begin{figure*}[t]
  \centering

  \begin{subfigure}[t]{0.32\textwidth}
    \centering
    \includegraphics[width=\linewidth]{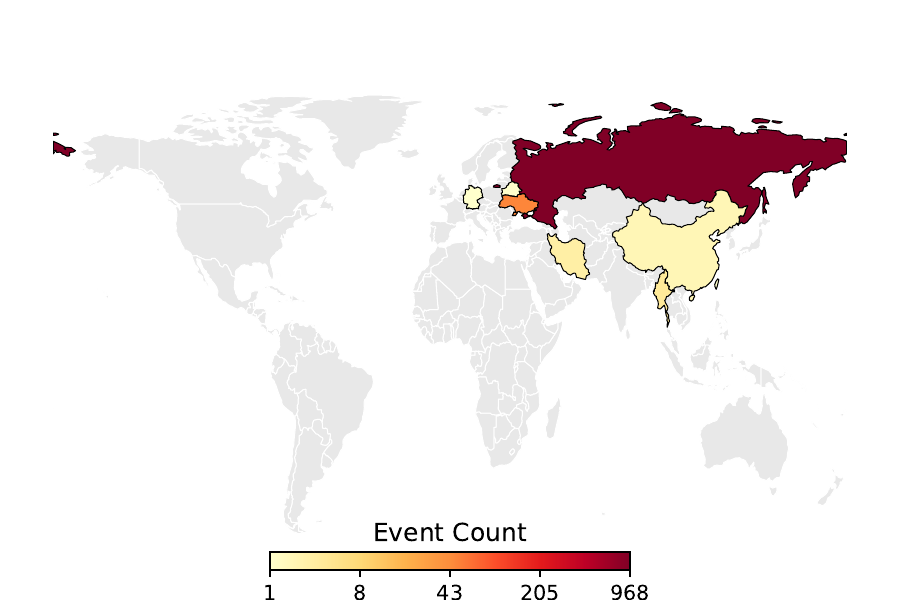}
    \caption{Russia as Intermediate Target}
    \label{fig:russia-div}
  \end{subfigure}\hfill
  \begin{subfigure}[t]{0.32\textwidth}
    \centering
    \includegraphics[width=\linewidth]{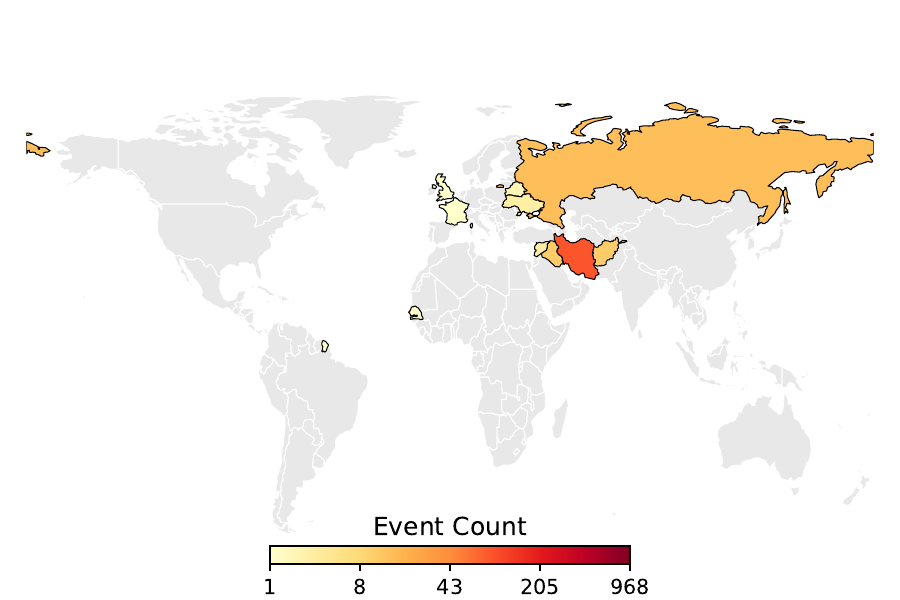}
    \caption{Iran as Intermediate Target}
    \label{fig:iran-div}
  \end{subfigure}\hfill
  \begin{subfigure}[t]{0.32\textwidth}
    \centering
    \includegraphics[width=\linewidth]{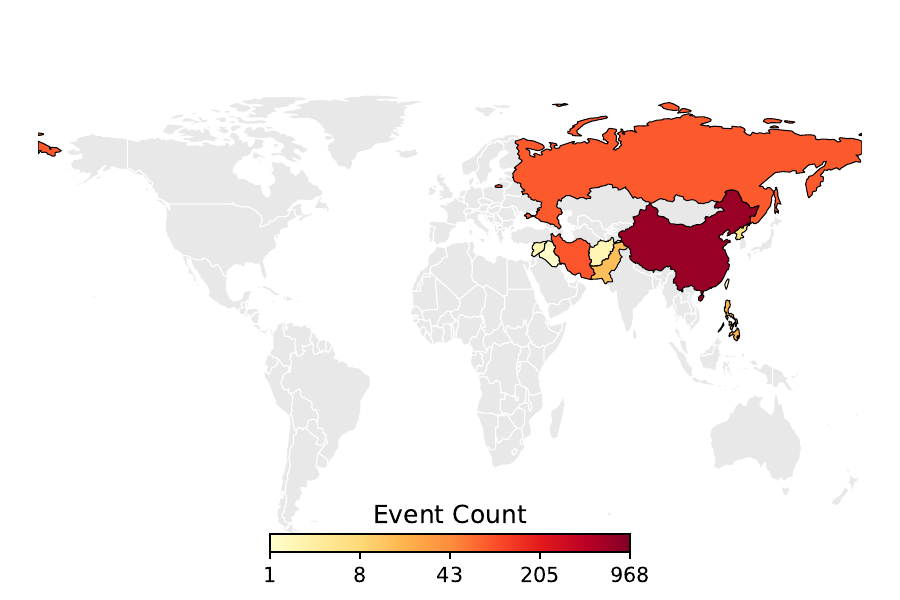}
    \caption{China as Intermediate Target}
    \label{fig:china-div}
  \end{subfigure}

  \caption{Intermediate Target Networks when Russia, Iran, and China as Intermediate Target, cases up to March 2025.}
  \label{fig:intermediate-networks}
\end{figure*}

\textbf{Internal Supply Chains}: China, Russia, and Iran each appear among the top members of their own Intermediate and Final Target Networks. Table~\ref{tab:top3_networks} summarizes this pattern. China and Russia occupy the top positions in both networks, while Iran ranks second in its Final Target Network but first in its Intermediate Target Network. This indicates that the U.S. most often identifies violations and imposes restrictions within domestic supply chains, where states channel restricted goods through domestic procurement and distribution circuits. The high volume of detected internal links also reflects a country's attempt to limit exposure to outside enforcement and pursue self-reliance. Such behaviors also make the referenced country itself the largest sanction buster. Figures~\ref{fig:intermediate-networks} and~\ref{fig:final-networks} also visualize this pattern, with each country’s own territory shaded darkest, indicating the highest number of sanction events related to itself.

\textbf{Tripartite Sanctions-Busting Networks}:
Although smaller in scale, the Intermediate Target Networks (INs) reveal a tightly interlinked structure among China, Russia, and Iran. As shown in Table~\ref{tab:top3_networks}, each country appears among the top three members of the others’ INs because recurring events bind them together. China frequently channels technology to Russia and Iran, while Russia provides routes back to China and Iran. Together, these links form a mutually reinforcing network that enables each state to maintain access to restricted technologies—reflecting shared geopolitical interests and coordinated efforts to counter U.S. sanctions.

\textbf{Network Structural Asymmetry}: A comparison of the corresponding panels between Figures~\ref{fig:intermediate-networks} and~\ref{fig:final-networks} (e.g., 7a vs.~8a) reveals clear asymmetries between Intermediate Target Networks (IN) and Final Target Networks (FN). FNs in Figure~\ref{fig:final-networks} span a geographically more diffuse set of countries across North America, Europe, and Asia, reflecting diversified procurement channels.  By contrast, the INs in Figure~\ref{fig:intermediate-networks} are more limited and concentrated geographically, involving only a few states. The different patterns reflect the globalized nature of procurement: strategic rivals diversify their supplier networks to secure advanced technologies, while intermediate targets profit from their sanction-busting business. Such dispersion makes U.S. sanctions harder to enforce, as removing a single supplier rarely disrupts the whole procurement network.

\textbf{Conclusion 3}: Subgraph analysis provides a macroscopic perspective unattainable through node- or edge-level metrics alone. Whereas RSI quantifies role and motifs decode strategic patterns, mapping the entire sanction network associated with a country makes overarching strategies empirically visible. It transforms thousands of discrete designations into coherent structural maps, exposing both the global reach of final target networks and the concentrated state linkages underpinning the intermediate target networks.

\section{Related Work}

\textbf{Temporal Bipartite Graph and Motif Analysis}: Temporal bipartite graphs model evolving relations between two distinct sets. Prior work has explored reachability and path queries in such graphs~\cite{10.14778/3467861.3467873}, motif structures like butterflies in static bipartite networks~\cite{10.14778/3725688.3725693}, and efficient enumeration of temporal motifs in general graphs~\cite{paranjape2017motifs}. Porter~\etal formalize motif statistics through the Temporal Activity State Block Model (TASBM)~\cite{10.1145/3487553.3524669}. Beyond algorithmic advances, motif analysis has revealed structural and behavioral dynamics across domains—from uncovering hidden transaction patterns in cryptocurrency~\cite{arnold2024insights}, to capturing information flows in mobile communication~\cite{kovanen2013temporal}, and illustrating hierarchical structures in organizational email networks~\cite{barnes2024temporal}. Our work differs in that it applies temporal motifs to sanction enforcement and export-control dynamics.

\textbf{Event-Based Modeling in Geopolitical Analysis}: Political science and computational social science have developed structured event datasets to quantify international dynamics. ICEWS~\cite{boschee2015icews} and GDELT~\cite{leetaru2013gdelt} extract events such as conflicts, protests, and sanctions from text, creating machine-readable databases. These resources support both macro-level indicator estimation~\cite{voukelatou2020estimating} and conflict case studies~\cite{keertipati2014multi}. While powerful, they largely operate at aggregate scales. Our approach instead leverages temporal motifs and subgraph structures to capture sanction dynamics with finer resolution and greater interpretability of strategies.

\section{Conclusion}
This paper introduced a temporal graph analytical framework that transforms the U.S. Entity List from a static compliance tool into a dynamic map of geopolitical strategy. By extracting event-level data from 25 years of Federal Register documents and modeling them as a temporal bipartite graph, we conducted a systematic analysis of sanction strategies at multiple levels of granularity. The Role Skew Index captures shifts in geopolitical alignment. The motif analysis detects and exposes the intensity of sanction campaigns. The subgraph enforcement networks reveal the broader ecosystems of US export control. 

Beyond these domain-specific insights, we provide a general methodology for analyzing regulatory data as temporal networks. Future work could extend this approach to other similar text documents, integrate trade and financial networks, or develop predictive models of sanction escalation. This study demonstrates that graph-based methods can decompose complex statecraft into measurable structures, offering new opportunities for computational research in social sciences.

\bibliographystyle{ACM-Reference-Format}
\bibliography{ref}

\appendix
\section{Comparison with Existing Datasets}

\textbf{Sanction Dataset}:
Two widely used country-level datasets on economic sanctions are the Threat and Imposition of Economic Sanctions (TIES) dataset and the Global Sanctions Database (GSDB). The TIES dataset~\cite{morgan2014threat} covers the 1945-2005 period. For each sanction episode, it distinguishes between threats and actual impositions, documenting the start and end dates, sender and target countries, and the type of sanction imposed, such as trade, financial, or military.  The Global Sanctions Database (GSDB)~\cite{FELBERMAYR2020103561} covers the 1950-2023 period and includes only imposed sanctions, excluding threats. It provides more detailed disaggregation by sanction type (trade, finance, export controls, or others) and further classifies trade sanctions into export, import, and bilateral trade restrictions. The dataset also codes political objectives (e.g., regime change, human rights, or territorial disputes), outcomes, and whether each case is unilateral or multilateral. 

\textbf{Entity List Dataset}:
Two main sources provide Entity List data. The Bureau of Industry and Security (BIS)~\cite{entity_list} publishes the current list as a snapshot of active designations. OpenSanctions~\cite{OpenSanctions} aggregates over 280 datasets, including the Entity List as part of the U.S. Trade Consolidated Screening List. Both data sources contain essential information about each entity, including appearance dates and recent changes. 

Our dataset differs in two key ways, making it more suitable for historical and strategic analysis. First, we capture every \texttt{add}, \texttt{modify}, and \texttt{remove} action as a distinct event, enabling the reconstruction of the list at any point in time and the analysis of sanction life cycles. Second, we introduce two critical attributes:

\begin{itemize}
\item Reason: the official justification for each listing, clarifying the specific policy concern (e.g., proliferation, human rights, or military end-use).
\item Distinguishing Intermediate and Final Target Countries: this allows us to turn the list of entities into a map of strategic relations.
\end{itemize}

Together, these features provide the first event-based view of the Entity List, enabling a systematic longitudinal analysis of U.S. detections and restrictions that is not possible with existing sources.

\section{Parsing Examples and LLM Configuration}
\label{sec:appendix_llm}
We present a representative example from a Federal Register notice alongside the LLM-parsed output.

\noindent\textbf{Sanction Case Example~\cite{sanction_example}}: The ERC determined to add the following entities to the entity list: 
\textit{Asia Pacific Links Ltd.} under the destination of China; 
\textit{Evolog Oy, Luminor Oy, and Siberica Oy} under the destination of Finland; 
\textit{ICW-Industrial Components Weirich} under the destination of Germany; and 
\textit{Device Consulting, Grant Instrument, SMT-iLogic, and Streloy} under the destination of Russia. 
These entities have been implicated in a conspiracy to violate U.S. export controls, including a scheme to supply the Special Technology Center, an entity on the BIS Entity List, with components to make unmanned aerial vehicles (UAVs) for Russia's Main Intelligence Directorate of the General Staff (GRU).

\noindent\textbf{Prompt Example}:
\begin{lstlisting}[basicstyle=\ttfamily\small,breaklines=true,frame=single]
SYSTEM ROLE:
You are a data analyst with knowledge in export-control regulations. Think step-by-step internally, but output only the final JSON content.
TASK:
1. Retrieve the best-matching XML record(s) for the entity_name or any alias.
2. Read the XML carefully and extract every field required by the pre-loaded JSON schema.
3. Extract or infer the reason for the queried entity to be listed.
4. Populate final_target_countries with the following rules: [see Section 3.2]
USER INPUTS:
entity_name: "ICW-Industrial Components Weirich"
entity_aliases: "None"
\end{lstlisting}

\noindent\textbf{Model Configuration}:
\begin{lstlisting}[basicstyle=\ttfamily\small,breaklines=true,frame=single]
model: gpt-4.1-mini
text.format: json_schema
temperature: 0.30
top_p: 1.00
max_output_tokens: 2048
store: true
json_schema: [See output example below]
\end{lstlisting}

\noindent\textbf{Model Example Output}:
\begin{lstlisting}[basicstyle=\ttfamily\small,breaklines=true,frame=single]
{
"entity_name": 
"ICW-Industrial Components Weirich",

"reasons_for_listing": 
"Conspiracy to violate U.S. export controls, 
including a scheme to supply the Special 
Technology Center with components to produce 
unmanned aerial vehicles (UAVs) for
Russia's Main Intelligence Directorate (GRU).",

"final_target_countries": ["Russia"]
}
\end{lstlisting}

Note that each API call in our pipeline queries exactly one entity that has already been extracted from the notice’s structured table (including names, aliases, addresses, and intermediate country). The LLM is not used for entity discovery or address parsing. Instead, it is limited to parsing the \texttt{reasons\_for\_listing} and \texttt{final\_target\_countries} fields. Document retrieval is supported via the file search tool on OpenAI’s hosting platform. Since each sanction case mentioning an entity is uniquely contained within a single document, we construct a per-document vector store. This design constrains the search space to only a small number of chunks, improving both precision and efficiency.

\textbf{LLM Parsing Errors}: 
We validated our LLM-assisted parsing on a random 10\% sample of 480 sanction events. Two researchers independently coded the \textit{final target country} and \textit{reason for listing} fields in the original Federal Register notices and compared their results with the LLM output. Agreement reached 96.25\% for \textit{final target country} and 97.50\% for \textit{reason for listing}.

Errors in identifying final targets (18/240 cases) fall into two types. First, the model sometimes misattributes the target to a country explicitly mentioned rather than the implied adversary—e.g., labeling Ukraine instead of Russia when the text refers to ``the situation in Ukraine.'' Second, some notices describe multi-country procurement networks without clearly naming the ultimate recipient, mixing actions and locations in long sentences that obscure the final target. For designation reasons (12/480 cases), errors mainly involve (1) overgeneralization, where specific items like ``photolithography machines'' or ``integrated circuits'' are simplified to ``semiconductors,'' and (2) incomplete extraction, when the shared justification appears separately from entity names and is therefore missed. Future work will refine prompts and contextual retrieval to improve accuracy in these complex and multi-country cases.

\begin{table}[htbp]
\centering
\caption{Mapping Between Clusters and Reason Categories}
\label{tab:cluster_reason_map}
\setlength{\tabcolsep}{4pt}
\begin{tabular}{@{}ll@{}}
\toprule
\textbf{Cluster ID} & \textbf{Category} \\
\midrule
0  & WMD Programs \& Delivery Systems \\
1  & -- \\
2  & Human Rights \& Surveillance \\
3  & Evasion \& Supply-Chain Facilitation \\
4  & Conventional Military \& Defense \\
5  & Semiconductors \& Advanced Computing \\
\midrule
\multirow{3}{*}{\makecell{Additional\\Categories}} & Energy \& Critical Infrastructure \\
                            & Cyber \& Secure Communications \\
                            & Other / Unknown \\
\bottomrule
\end{tabular}
\end{table}

\section{Categorization of Listing Reasons}
\label{sec:appendix_cat}
\begin{figure}[htbp]
    \centering
    \includegraphics[width=0.8\linewidth]{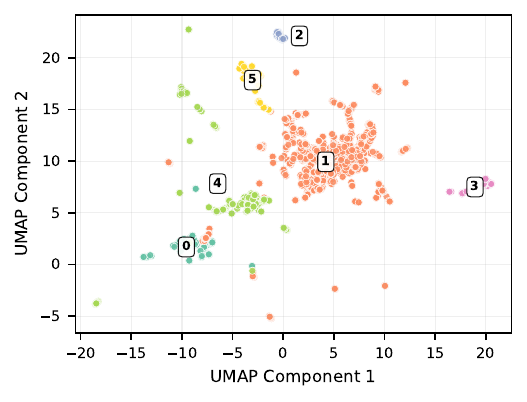}
    \caption{Clustering of Extracted Reason Statements}
    \label{fig:reason_clustering}
\end{figure}

The extracted \textit{reasons for listing} are long narrative sentences that often contain entity names, legal boilerplate, and policy phrases. We first condense these into short phrases that capture the controlled activity and its capability area by removing non-informative text. We then generate vector embeddings using OpenAI’s \texttt{text-embedding-small} model. An initial $k$-means clustering produces six clusters (Figure~\ref{fig:reason_clustering}) with a silhouette score of 0.61, which we further refine into eight interpretable categories labeled by common keywords and reason descriptions. Table~\ref{tab:cluster_reason_map} shows the mapping between clusters and categories.
Cluster~1 is examined in detail and further divided, with its subgroups mainly assigned labels consistent with Clusters~3 and~4. Additional labels are created for a few unique cases that do not align with the main categories.

\begin{table*}[htbp]
\centering
\caption{Term-level motif peak statistics under the case-shuffled null.}
\label{tab:term_null}
\begin{tabular}{@{}l
                >{\raggedright\arraybackslash}p{3.2cm}
                rrrrc@{}}
\toprule
Country & Term (Peak Year) & \multicolumn{1}{c}{Obs Count} & 
\multicolumn{1}{c}{Null Mean} & 
\multicolumn{1}{c}{Null Std} & 
\multicolumn{1}{c}{$p_{c,\tau}$} & Sig \\
\midrule
Russia 
 & Biden (2024) & 151{,}781.0 & 54{,}210.1 & 3{,}917.48 & 0.0010 & $^{\ast\ast\ast}$ \\\hline

\multirow{4}{*}{Iran} 
 & Bush (2008)  & 667.0   & 94.6   & 48.50   & 0.0010 & $^{\ast\ast\ast}$ \\
  & Obama (2011) & 5{,}124.0  & 645.3  & 187.59 & 0.0010 & $^{\ast\ast\ast}$ \\
 & Obama (2016) & 7{,}075.0  & 2{,}549.0  & 487.61  & 0.0010 & $^{\ast\ast\ast}$ \\
 & Trump (2020) & 7{,}394.0  & 4{,}080.4  & 642.14  & 0.0010 & $^{\ast\ast\ast}$ \\
 & Biden (2022) & 4{,}006.0  & 9{,}704.4  & 1{,}257.46 & 1.0000 & n.s. \\\hline

\multirow{2}{*}{China}
 & Trump (2020) & 17{,}353.0 & 4{,}157.3  & 721.79  & 0.0010 & $^{\ast\ast\ast}$ \\
 & Biden (2024) & 45{,}604.0 & 23{,}395.7 & 2{,}245.93 & 0.0010 & $^{\ast\ast\ast}$ \\
\bottomrule
\end{tabular}

\footnotesize
Notes:  All results are based on $R = 1000$ permutations.``n.s'' denotes no significant peak ($p_{c,\tau} \ge 0.005$). 
\end{table*}

\section{Statistical Significance for Motif Counts}
\label{sec:appendix_null}
To evaluate whether observed motif bursts reflect genuine coordinated sanction campaigns rather than random temporal coincidence, we construct a null model that retains the structural properties of sanction events but randomizes their temporal patterns.  

The simplest null model would be a global time-shuffled model that preserves each edge’s endpoints and duration but randomly redistributes all start times across the full observation window. This model can test if motif bursts are merely artifacts of random timing. However, this model erases the natural variation in annual event activities. The null baseline becomes unrealistically flat, making any peak appear more exceptional than it truly is. In years of heightened enforcement or geopolitical escalation, the number of sanctioned entities increases dramatically. To test whether motif bursts remain statistically meaningful even in ``busy'' years, we therefore adopt a stricter null that preserves both case structure and yearly intensity.

\textbf{Case-shuffled null model}: Let $\mathcal{P} = \{(u(e), v(e)) : e \in E \}$ be the set of all observed country pairs. 
For each year $y$, let $E_y$ denote the set of events whose start time $t^{\mathrm{add}}(e)$ falls in year $y$. To generate one realization of the null model:
\begin{itemize}
  \item \textbf{Preserve year-level activity:} Keep the number of events $|E_y|$ in each year identical to the observed data.
  \item \textbf{Preserve durations:} each event $e_k$ retains its original duration $\Delta(e_k) = t^{\mathrm{rm}}(e_k) - t^{\mathrm{add}}(e_k)$.
  \item \textbf{Reassign pairs:} Permutate the entire set $\mathcal{P}$ and reassign these country pairs to the original events in order, maintaining each event’s original year and start time.
\end{itemize}

In essence, the question is: even in years of intense sanction activity, would random reallocation of pairs produce motif peaks as strong as those observed?

\textbf{Permutation test and $p$ value}: For a given country $c$ and presidential term $\tau$, let $S_y(c^{\mathrm{fin}})$ denote the annual count of final-target motif and $S_y(c^{\mathrm{int}})$ denote the annual count of intermediate target motif. We define the observed term-level statistic as the dominant motif's peak within the term:

\[
T^{\mathrm{obs}}_{c,\tau}
= 
\max\!\Bigl\{
\max_{y \in \tau} S_y(c^{\mathrm{fin}}),
\ \max_{y \in \tau} S_y(c^{\mathrm{int}})
\Bigr\}
\]

For each of $R$ ($r = 1,\dots,R$) permutations, we recompute the motif count under the pair-reassignment null and obtain:

\[
T^{(r)}_{c,\tau}
=
\max\!\Bigl\{
\max_{y \in \tau} S^{(r)}_{y}(c^{\mathrm{fin}}),
\ \max_{y \in \tau} S^{(r)}_{y}(c^{\mathrm{int}})
\Bigr\}
\]

The one-sided permutation $p$-value is then given by:
\[
p_{c,\tau}
=
\frac{1 + \#\{\, r : T^{(r)}_{c,\tau} \ge T^{\mathrm{obs}}_{c,\tau} \,\}}{R + 1}
\]

\begin{table*}[htbp]
\centering
\caption{Members in China's Intermediate and Final Target Networks (Cases up to March 2025)}
\label{tab:china_member}
\begin{tabular}{p{0.47\linewidth} p{0.47\linewidth}}
\toprule
\multicolumn{2}{c}{\textbf{Intermeidate Target Network}} \\
\midrule
China (694) & Iran (87) \\
Russia (83) & Philippines (19) \\
Pakistan (15) & North Korea (6) \\
Macau (4) & Hong Kong (3) \\
Afghanistan (2) & Syria (2) \\
Singapore (2) & Iraq (1) \\
Taiwan (1) & \\
\midrule
\multicolumn{2}{c}{\textbf{Final Target Network}} \\
\midrule
China (694) & Hong Kong (75) \\
Singapore (12) & South Africa (10) \\
United Arab Emirates (10) & United Kingdom (7) \\
Germany (4) & Thailand (4) \\
Netherlands (4) & Belgium (4) \\
Mexico (3) & Japan (3) \\
Taiwan (3) & Qatar (3) \\
Pakistan (3) & France (3) \\
Russia (3) & Switzerland (2)\\
India (2) & Cyprus (2) \\
Brazil (2) & Chile (2) \\
British Virgin Islands (2) & Italy (2) \\
Egypt (1) & Denmark (1) \\
Cuba (1) & Cayman Islands (1) \\
Belarus (1) & Argentina (1) \\
Australia (1) & Bahrain (1) \\
Costa Rica (1) & Canada (1) \\
Bolivia (1) & Israel (1) \\
Kenya (1) & Peru (1) \\
Panama (1) & New Zealand (1) \\
Myanmar (1) & Morocco (1) \\
Laos (1) & Madagascar (1) \\
Malaysia (1) & Jamaica (1) \\
South Korea (1) & Portugal (1) \\
Romania (1) & Sri Lanka (1) \\
Sweden (1) & Turkey (1) \\
Vietnam (1) & \\

\bottomrule
\end{tabular}
\end{table*}

This $p_{c,\tau}$ represents the probability that, under the case-shuffled null model, a motif peak as large as or larger than the observed one would occur within the same presidential term. Small values of $p_{c,\tau}$ therefore indicate that even after accounting for high annual activity, the observed motif burst reflects coordinated sanctioning behavior rather than random pair alignments.

Table~\ref{tab:term_null} reports the observed motif count peaks and their significance under the pair-reassignment null. Across all three countries, most peaks remain highly significant ($p_{c,\tau}<0.005$). Only Iran’s 2022 shows no significant peak. However, this is consistent with the declining motif count observed during that period in Figure~\ref{fig:three_county_motif_groups}.

\section{Subgraph Examples}
\label{sec:appendix_net}
To illustrate the composition of each country's network, Table~\ref{tab:china_member} lists all members appearing in China's Intermediate and Final Target Networks. Each entry is labeled with the number of sanction cases in which the country appears. The Final Target Network spans a broad set of countries, while the Intermediate Target Network is much smaller. 

\end{document}